\documentclass{kluwer}    
\usepackage{graphicx}
\newdisplay{guess}{Conjecture}

\begin{document}                                                                                   
\begin{article}
\begin{opening}         
\title{X-ray Observations of Binary and Single Wolf-Rayet 
       Stars with XMM-Newton and Chandra} 
\author{Stephen \surname{Skinner}}
\institute{CASA, 389 UCB, Univ. of Colorado, Boulder, CO 80309 USA}
\author{Manuel   \surname{G\"{u}del}}
\institute{Paul Scherrer Inst., W\"{u}renlingen und Villigen, CH-5232
           Villigen PSI, Switzerland}
\author{Werner \surname{Schmutz}}
\institute{Physik-Met. Observatorium Davos (PMOD), Dorfstrasse 33, 
           CH-7260 Davos Dorf, Switzerland}
\author{Svetozar \surname{Zhekov}}
\institute{Space Research Inst., Moskovska Str. 6, Sofia 1000,
           Bulgaria, and JILA, 440 UCB, Univ. of Colorado,
           Boulder, CO 80309 USA}

\runningauthor{Stephen Skinner}
\runningtitle{X-ray Observations of WR Stars}
\date{July 15, 2005}

\begin{abstract}
We present an overview of recent X-ray observations of Wolf-Rayet (WR)
stars with {\em XMM-Newton} and {\em Chandra}. These observations are aimed 
at determining the differences in X-ray properties between massive 
WR $+$ OB binary systems and putatively single WR stars.
A new {\em XMM} spectrum of the nearby WN8 $+$ OB binary WR 147 shows hard
absorbed X-ray emission (including the Fe K$\alpha$ line complex),
characteristic of colliding wind shock sources. In contrast, sensitive
observations of four of the closest known single WC (carbon-rich) WR
stars have yielded only non-detections. These results tentatively suggest
that single WC stars are X-ray quiet. The presence of a companion may thus be
an essential factor in elevating the X-ray emission of 
WC $+$ OB stars to detectable
levels.
\end{abstract}

\end{opening}           

\section{Wolf-Rayet Stars}  

Wolf-Rayet (WR) stars are luminous objects that are losing mass at
very high rates $\dot{\rm M}$ $\sim$ 10$^{-5}$ -
10$^{-4}$ M$_{\odot}$ yr$^{-1}$. WR stars are the evolutionary
descendants of massive O-type stars and are in advanced
nuclear burning stages, approaching the end of their lives
as supernovae. They are broadly classified 
as either WN or WC subtypes according to whether N or C
lines dominate the optical spectra. WC stars show chemical evidence
of He burning and are thought to be more evolved than WN
stars. The powerful winds of WR stars  enrich the ISM with 
heavy elements that will ultimately be recycled into future
generations of stars. X-ray studies of WR stars provide crucial
information on physical conditions in their winds and outer
atmospheres. Detailed comparisons of X-ray properties with shock
models can be used to infer information on mass-loss properties.
High-resolution X-ray grating spectra also offer the potential
of measuring chemical abundances in their metal-rich winds.

\begin{figure}
\centerline{\includegraphics[width=1.7in]{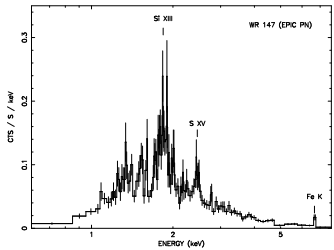}}
\caption{X-ray spectrum of the composite WR 147 N$+$S
system binned to a minimum of 15 counts per bin. Data
are from the {\em XMM} EPIC-PN camera using the MED
optical blocking filter. Prominent emission lines are
identified.}
\end{figure}

\section{X-rays from Wolf-Rayet Binaries}

Strong X-ray emission has been detected from several
massive WR $+$ OB binary systems (e.g. 
Skinner et al.  2001). It is
believed that the X-rays originate mainly in a colliding
wind (CW) shock region between the two stars. However,
other mechanisms may be involved. Intrinsic emission from
the stars themselves may be present, and processes such as
radiative recombination far out in the winds can give rise 
to X-ray spectral lines (Schild et al. 2004).

WR 147 is an excellent target for testing the predictions
of CW models because it lies  nearby at a 
distance of 0.63 kpc and its stellar properties 
are  relatively well-known. {\em HST} observations show that the
system consists of WN8 star lying $\approx$0.62$''$ south of a 
O5-7 star (L\'{e}pine et al.  2001), but other studies assign 
a later B0 type for the companion. Several high-resolution radio and
IR observations indicate that a CW shock is present near the surface
of the OB star. Strong X-ray emission was detected by {\em ASCA}
(Skinner et al. 1999), and {\em Chandra} shows the       emission
is extended on arcsecond scales (Pittard et al. 2002).

Our sensitive 22 ksec {\em XMM} observation of WR 147 reveals a strongly
absorbed  X-ray spectrum and high-temperature emission lines (Fig. 1).
The  Fe K$\alpha$ complex at 6.67 keV is detected for 
the first time in this system. Acceptable spectral fits with
two-temperature thin plasma models give an absorption column
density  log N$_{\rm H}$ = 22.4 cm$^{-2}$ with plasma temperatures
kT$_{cool}$ $\approx$ 0.8 keV (9 MK) and kT$_{hot}$ $\approx$ 
3 - 4 keV (35 - 46 MK). The derived absorption is consistent
with estimates based on the known large visual extinction 
A$_{V}$ = 11.5 mag. Detailed modeling of the X-ray spectrum is
underway to determine if the observed X-ray properties are consistent
with CW shock predictions based on currently accepted mass-loss
parameters of WR 147. 

\begin{table} %
\begin{tabular*}{\maxfloatwidth}{lllrr}                                        
\hline
Star    & Sp. type & dist. & log L$_{\rm X}$ & log[L$_{\rm X}$/L$_{bol}$] \\
        &          & (kpc) & (erg/s)         &                            \\
WR 147  & WN8 + OB & 0.63  & 32.60           & $-$6.7          \\
WR 5    & WC6      & 1.91  & $\leq$30.97     & $\leq$$-$7.7    \\
WR 57   & WC8      & 2.37  & $\leq$30.95     & $\leq$$-$8.1    \\
WR 90   & WC7      & 1.64  & $\leq$30.29     & $\leq$$-$8.7    \\
WR 135  & WC8      & 1.74  & $\leq$29.82     & $\leq$$-$9.1    \\

\hline
\end{tabular*}
\caption[]{X-ray Data for WR Stars} 
{\em Table Note}: Upper limits on the unabsorbed X-ray luminosity
in the 0.5 -- 7 keV band are from the {\em PIMMS} mission
simulator, assuming a 1 keV thermal plasma with mean extinctions and
distances from van der Hucht (2001).  The {\em XMM} upper limits for
WR 5, 57, and 90 are for 68\% encircled energy (15$''$ extraction radii)
based on mean background  MOS count rates (MED filter). Time intervals
during  background flares were excluded; the WR5 observation
was severely affected by high background.  
The {\em Chandra} upper limit for WR 135 assumes a 5 count 
detection threshold in the ACIS-S detector. 
L$_{bol}$ values are
from Koesterke \& Hamann (1995); the value for WR 90 is an
average for WC7 stars.
\end{table}

\section{X-rays from Single Wolf-Rayet Stars}

In contrast to WR $+$ OB binaries, pointed X-ray observations of
single WR stars are almost non-existent and little is known about
their X-ray properties. Characterizing their emission is important
for interpreting the X-ray spectra of  WR $+$ OB systems,
whose components are usually too closely spaced to be resolved in
X-rays. Single OB stars have been detected in X-rays and 
almost always show soft emission at characteristic temperatures
kT $<$ 1 keV  (Bergh\"{o}fer et al. 1997). Theoretical models generally
attribute the soft emission to shocks that are formed in their
radiatively-driven winds, but other processes involving magnetic
phenomena may also be at work in young OB stars.

Given that single OB stars are X-ray emitters, one expects that  
single WR stars should be as well. To investigate this question,
we are obtaining sensitive X-ray observations of selected WR stars.
To date, we have observed four of the closest known single WC stars (Table 1),
but single WN stars have not yet been studied. 
Surprisingly, none of the four WC stars were detected! 
Our most stringent upper limit is based on a 
20 ksec {\em Chandra} observation of WR 135, which detected only
one X-ray count in a nominal source extraction region centered on the star's
optical position (consistent with background predictions).
This implies a remarkably low intrinsic X-ray luminosity
L$_{\rm X}$ $\leq$ 10$^{29.1}$ ergs s$^{-1}$, but we quote a more conservative
upper limit in Table 1 based on a 
nominal 5-count {\em Chandra} detection threshold.
The L$_{\rm X}$/L$_{bol}$ value of WR 135 is at least two orders of 
magnitude below what is typically seen for O stars (Fig 2.). These
initial results suggest that single WC stars may be X-ray quiet,
but additional observations are needed to enlarge the sample. The
reason for the apparent suppression of their X-ray emission is not yet
known, but X-ray absorption in their dense metal-rich winds may be 
a factor.

\begin{figure}
\centerline{\includegraphics[width=12pc]{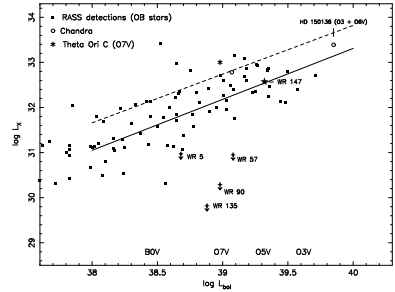}}
\caption{X-ray versus bolometric luminosity for O stars detected
in RASS (solid squares), with Wolf-Rayet stars discussed in the
text overlaid for comparison. The solid line is a regression fit
for O stars based on RASS detections and non-detections
(Bergh\"{o}fer et al. 1997). The dashed line is a similar 
regression fit based on  {\em Einstein} data for O stars (not shown).
The two circles are based on {\em Chandra} data for
the O stars HD 150135 (O6.5V) and HD 150136 (O3 $+$ O6V) from
Skinner et al. (2005). The downward arrows denote X-ray upper
limits for undetected WC stars.}
\end{figure}



\acknowledgements
This work was supported by NASA grants NNG05GA10G, NNG05GB48G, and GO 5003-X.



\end{article}
\end{document}